\documentclass[journal]{IEEEtran}
\usepackage{cite}
\usepackage{amsmath,amssymb,amsfonts}
\usepackage{graphicx}
\usepackage{textcomp}
\usepackage{xcolor}
\usepackage{algpseudocode}
\usepackage{algorithm}
\usepackage[hidelinks]{hyperref}
\def\BibTeX{{\rm B\kern-.05em{\sc i\kern-.025em b}\kern-.08em
    T\kern-.1667em\lower.7ex\hbox{E}\kern-.125emX}}

\newcommand{\bmat}[1]{\begin{bmatrix}#1\end{bmatrix}}

\begin{document}
\title{Linearity, Time Invariance, and Passivity of a Novice Person in Human Teleoperation}

\author{\IEEEauthorblockN{David Black,}
\and
\IEEEauthorblockN{Septimiu Salcudean}
}

\maketitle

\begin{abstract}
Low-cost teleguidance of medical procedures is becoming essential to provide healthcare to remote and underserved communities. Human teleoperation is a promising new method for guiding a novice person with relatively high precision and efficiency through a mixed reality (MR) interface. Prior work has shown that the novice, or ``follower", can reliably track the MR input with performance not unlike a telerobotic system. As a consequence, it is of interest to understand and control the follower's dynamics to optimize the system performance and permit stable and transparent bilateral teleoperation. To this end, linearity, time-invariance, inter-axis coupling, and passivity are important in teleoperation and controller design. This paper therefore explores these effects with regard to the follower person in human teleoperation. It is demonstrated through modeling and experiments that the follower can indeed be treated as approximately linear and time invariant, with little coupling and a large excess of passivity at practical frequencies. Furthermore, a stochastic model of the follower dynamics is derived. These results will permit controller design and analysis to improve the performance of human teleoperation.
\end{abstract}


\section{Introduction}
Human teleoperation enables remote control of medical procedures such as ultrasound exams in isolated or under-resourced communities \cite{yeung2024}, as well as remote maintenance, hand-over-hand teaching, and many other applications. Compared to existing video teleguidance systems, for example as used in tele-ultrasound, it offers significantly improved efficiency and precision \cite{black2023hci}, while it is more practical and accessible than robotic teleoperation due to lower cost, size, and complexity, and increased flexibility \cite{black2025rob}.

\begin{figure*}[t]
    \centering
    \includegraphics[width=0.9\linewidth]{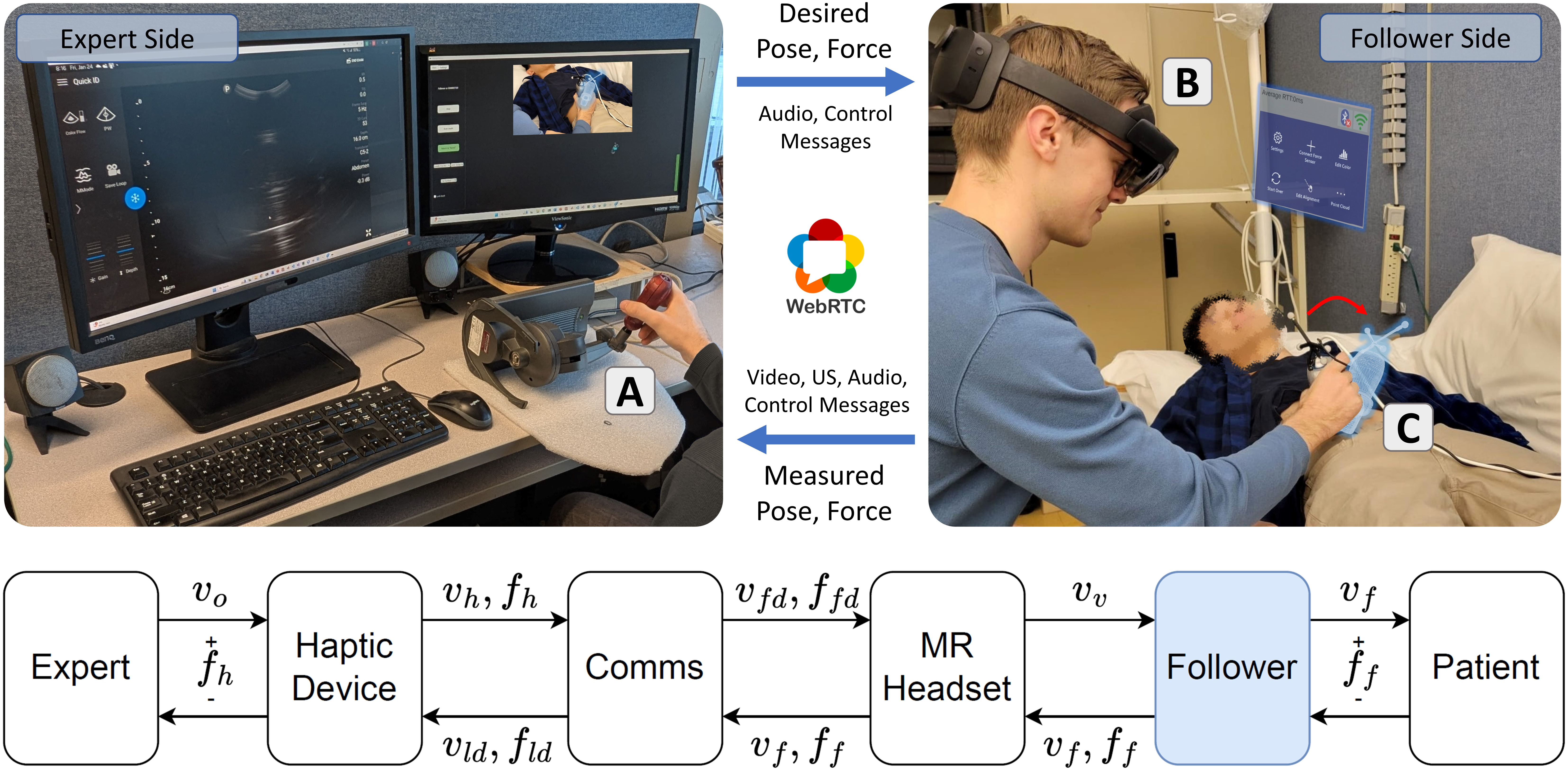}
    \caption{Overview diagram and corresponding two-port network representing the human teleoperation system, showing the follower block in blue. (A) the haptic device, (B) MR headset, (C) follower's instrumented tool and the virtual tool. The red arrow shows how the follower will move next to align the real tool to the virtual one. In this example, the system is used for teleultrasound.}
    \label{fig:twoPort}
\end{figure*}

In human teleoperation, shown in Fig. \ref{fig:twoPort}, a novice follower person aligns their tool with a 3D virtual one displayed to them through a mixed reality (MR) headset. The motion of the virtual tool is controlled in real time by a remote expert, who manipulates a 6 degree of freedom (DOF) haptic device and views a live stream of the ultrasound image \cite{black2024cag}. Prior work has shown through tracking tests as well as step and frequency response measurements that the dynamics of the follower person in attempting to align their real tool with the virtual one resemble a second-order linear system and the performance can be compared to telerobotic systems \cite{black2023ijcars,black2024tmrb}.

By tracking the follower's position and orientation (pose) \cite{black2024pose} and force \cite{black2024ft}, it is possible to develop a bilateral teleoperation system to shape the follower's behavior and provide haptic feedback to the expert \cite{black2024bilat}. To choose and design appropriate controllers, it is important to understand whether the follower's dynamics in response to an MR input can be treated as a linear, time invariant (LTI) system. As linear control theory is easier and more developed than nonlinear control \cite{slotine1991, vaidyanathan2016}, having an accurate linear model would greatly simplify the design and analysis of the controller. Moreover, most common methods assume a time invariant plant whereas time-varying systems require more careful and complex methods such as gain scheduling or adaptive control \cite{marino2003,byrnes1995}.

On first inspection, the human arm is a highly nonlinear system \cite{tolani1996}. It forms a complex kinematic chain akin to a serial robotic manipulator, and has elastic actuation. Moreover, unlike mechanical systems, people are susceptible to distraction, fatigue, and random variance in their performance. Thus, a nonlinear, time-varying model is perhaps most appropriate. However, robotic manipulators can be controlled through feedback linearization using computed torque \cite{an1987} or resolved-acceleration \cite{luh1980} methods. By feeding forward the dynamic model of the robot plus the desired trajectory wrapped in a PD controller, the robot's nonlinear dynamics are effectively canceled out. Assuming a perfect model, this leaves linear, uncoupled error dynamics for which the PD gains can be tuned~\cite{corke2011}. 

No robotic manipulator exists that can move with the same combination of speed, flexibility, and compliance of a human arm, so it may be that our own internal controllers also lead to LTI, uncoupled dynamics. Indeed some papers have modeled the human arm dynamics as a second order linear system given a desired position commanded by the central nervous system \cite{dolan1993}. However, the question of whether the human dynamics in response to MR pose inputs are LTI and/or coupled remains unanswered.

Importantly, the teleoperation system must also be made robust to time delays since there are communication lags and latency associated with the follower's response time \cite{black2024tmrb}. The problem of stable and transparent bilateral teleoperation in the presence of time delays has been very heavily studied \cite{hokayem2006, sheridan1993, muradore2016, deng2021, bolopion2013}. Several of the most common methods for delay-robust bilateral teleoperation rely on the concept of passivity to guarantee stability. These include wave variables \cite{niemeyer1991,aziminejad2008} and passivation of the communication channel \cite{anderson1988}, the time domain passivity approach (TDPA) \cite{hannaford2002,ryu2004teleop}, and many variations \cite{ryu2010,panzirsch2019, choi2022}.

Passivity can be understood in analogy to electrical circuits, where passive elements such as resistors, inductors, and capacitors may store or dissipate energy but not create it. Indeed, a teleoperation system, like a circuit, can be represented as a network of one and two port elements \cite{anderson1988}. An element that outputs no more energy than it receives as input is called passive. An element that dissipates energy is strictly passive, or dissipative. More generally, in controls, passivity is a characteristic of the input-output relationship and need not pertain to physical units of energy \cite{nuno2011}. 

Consider a system with input $\pmb{u}$ and output $\pmb{y}$ such that $\pmb{y}=T\pmb{u}$. The generalized instantaneous power, or energy flow into and out of the system is $p(t)=\pmb{y}(t)^\top\pmb{u}(t)$. The usual convention treats energy flow into the system as positive, while out is negative. The system is passive if and only if
\begin{equation}
    E(0)+\int_0^t\pmb{y}(\tau)^\top\pmb{u}(\tau)d\tau\geq 0
\end{equation}
That is, the total energy including the initial state is non-negative at all times. A strict inequality here would lead to strict passivity. Note, there is no requirement for $\pmb{y}^\top\pmb{u}$ to have physical units of power.

In teleoperation, the operator's hand and the robot's environment are 1-port terminations to the 2-port network and are commonly considered to be passive \cite{hogan1987}. Such a system is stable for any passive environment and operator blocks if and only if the structured singular value of the scattering operator, $S$, satisfies $\sup_\omega\mu_\Delta(j\omega)\leq1$ and is BIBO stable \cite{colgate1988}. It is straightforward to design passive controllers for the leader and follower robots. Thus, together, these factors make it relatively simple to guarantee the stability of the system by making the communication passive.

However, it has been shown that the environment's passivity may be limited \cite{willaert2009} or non-existent \cite{jazayeri2015}, for example during robotic surgery inside a beating heart. Moreover, unlike in the velocity-force domain, the operator's arm impedance may be active in some frequency ranges in the position-force domain \cite{shahbazi2018}, i.e. when $\pmb{u}$ is position, not velocity. Indeed, for anything in the system to move at all, it requires an active injection of energy, in much the same way that an isolated circuit containing only resistors will not spontaneously set up an electric current. The key to maintaining overall passivity is that this active component or shortage of passivity (SoP) is compensated by an excess of passivity (EoP) elsewhere in the system \cite{jazayeri2015}.

In any case, understanding the passivity of every component in a teleoperation system is essential to creating stable bilateral controllers. Beyond the work of Hogan \cite{hogan1987} and Shahbazi \cite{shahbazi2018} exploring the passivity of the operator's arm, others have investigated the energetic passivity of the human ankle \cite{lee2016}. However, it is unknown to what degree, if at all, the follower in a human teleoperation system can be considered passive.

This paper therefore sets out to understand practical control theoretic aspects of the dynamics of the human follower attempting to track a virtual tool in MR. Through modeling and empirical analysis, the following questions are answered:

Can the follower in human teleoperation be approximated as a(n)
\begin{itemize}
    \item Linear system?
    \item Time-invariant system?
    \item Uncoupled system? (i.e. no inter-axis coupling)
    \item Passive system?
\end{itemize}
The role of randomness in the follower response is additionally characterized.

\section{Methods}
Human teleoperation can be represented as shown in Fig. \ref{fig:twoPort} as a network of one and two port elements. This is similar to robotic teleoperation except that the follower robot is replaced by the MR headset and follower person blocks. Like a robot, the controller of the visual input on the MR headset can be designed to be passive. The goal of this paper is therefore to understand the last remaining unknown block: the follower.

Several assumptions have to be made to start the analysis. We assume the follower is physically able to track a virtual tool, mentally unimpaired and able to focus on the task, and committed to tracking the tool as well as possible. The latter ensures that they do not willfully diverge from the desired trajectory. It is further required that the frequency content of the input trajectory is low enough for a person to follow relatively comfortably, and that the trajectory is reachable - e.g. that the virtual tool does not move deep into the environment where the follower would have to apply excessive forces to match it. For any practical application of human teleoperation, these assumptions are necessary for the usability of the system, so they do not limit the analysis in any way.

In this case, the follower can be modeled as a mass-spring-damper connected to the virtual tool \cite{black2024bilat}. The output force is zero in free space, whereas in contact, the environment is modeled as a spring-damper. The environment spring-damper has no effect on the follower's motion, as they try exclusively to track the virtual probe. This is where the assumptions of the follower's willingness to track the virtual tool and the fact that the tool is not too far inside the environment become important. Thus, the follower can be written in state space as
\begin{align}\label{eqn:stateSpace}
    \pmb{\dot{x}}_f &= \bmat{\dot{x}_f\\\ddot{x}_f}=A\pmb{x}_f+B\pmb{u}\nonumber\\
    \pmb{y}_f &= \bmat{x_f\\f_f}=C\pmb{x}_f
\end{align}
Where
\begin{align}
    A&=\bmat{0&1\\-\dfrac{k_f}{m_f} & -\dfrac{b_f}{m_f}}\nonumber\\
    B&=\bmat{0&0\\\dfrac{k_f}{m_f} & \dfrac{b_f}{m_f}}\nonumber\\
    C&=\bmat{1&0\\k_p & b_p}
\end{align}
The input, $\pmb{u}$, is the position and velocity of the virtual tool. The state is the position and velocity of the follower's tool, and the output is the position and force of the follower tool, if interacting with the environment.

In a three-axis follower model (considering all positions, but not rotations), the matrices would instead be $A_3\in\mathbb{R}^{6\times6}$, $B_3\in\mathbb{R}^{6\times6}$, $C_3\in\mathbb{R}^{3\times6}$, and $D_3\in\mathbb{R}^{3\times6}$. If the positional axes have negligible coupling, we can consider one DOF at a time and simplify the remaining analysis to one dimension, as in Equation \ref{eqn:stateSpace}. This is tested in Section \ref{sec:axis}, as is the behavior in orientation space.

The above model clearly assumes linearity, and is tested in Section \ref{sec:linear}. Moreover, time variance could be captured by making the model parameters, $k_f$, $b_f$, and $m_f$ functions of time. Whether this is necessary is discussed in Section \ref{sec:ti}.

\subsection{Experiments}
\begin{figure}[h]
    \centering
    \includegraphics[width=\linewidth]{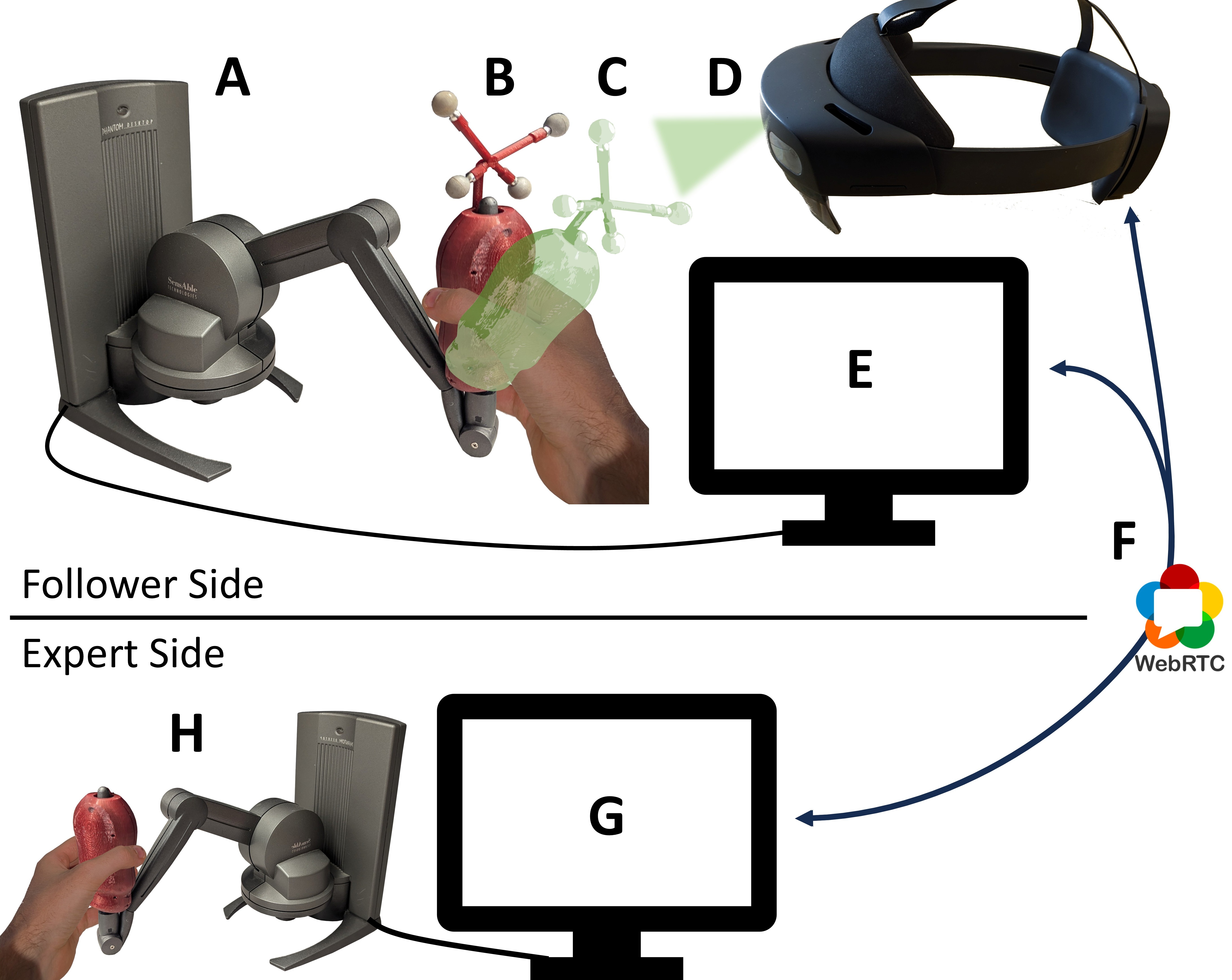}
    \caption{Diagram of the experimental system. (A) Follower haptic device. (B) Ultrasound probe-shaped handle with IR markers. (C) Virtual tool in the same shape. (D) Microsoft HoloLens 2 worn by the follower to render the virtual tool. (E) Follower-side computer to control the haptic device. (F) WebRTC communication between expert and follower PCs and the HoloLens. (G) Expert-side computer to generate trajectories and control their haptic device. (H) Expert haptic device (not used in these experiments).}
    \label{fig:system}
\end{figure}

To determine the linearity, time dependence, passivity, and inter-axis coupling of a follower person guided through MR, we carried out a range of experiments and theoretical analyses, the results of which are presented in the next section. For the experiments, a similar setup to that described in \cite{black2024bilat} was used, as shown in Fig. \ref{fig:system}. 

In short, two Touch X haptic devices (3D Systems, Rock Hill, SC) were used to track the motions of and apply forces to the follower and expert respectively. The follower's device had a 3D printed ultrasound probe shaped end effector with infrared (IR) reflective spheres. The follower also wore a Microsoft HoloLens 2 MR headset to view the virtual ultrasound probe. The headset was registered to the haptic device by tracking the IR markers \cite{black2024pose} for 1000 samples while moving the haptic device's handle slowly, and then computing the transformation matrix using the Kabsch-Umeyma algorithm \cite{lawrence2019}. In some of the tests, the follower's haptic device rendered a flat, horizontal virtual surface with stiffness $k_p$ and damping $b_p$ to simulate contact with an environment. 

The expert's haptic device motions were transmitted over a fast Web Real Time Communication (WebRTC) link to the follower PC as well as to the follower's HoloLens 2. After transformation with the registration matrix, the poses were displayed to the follower using a 3D virtual ultrasound probe displayed in the headset. The follower then aligned his or her haptic device with dummy ultrasound probe end effector to the virtual probe and attempted to follow it as accurately and quickly as possible. The expert's and follower's positions, orientations, and forces were recorded with timestamps at a rate of approximately 100 Hz. 
\begin{figure*}[t]
    \centering
    \includegraphics[width=\linewidth]{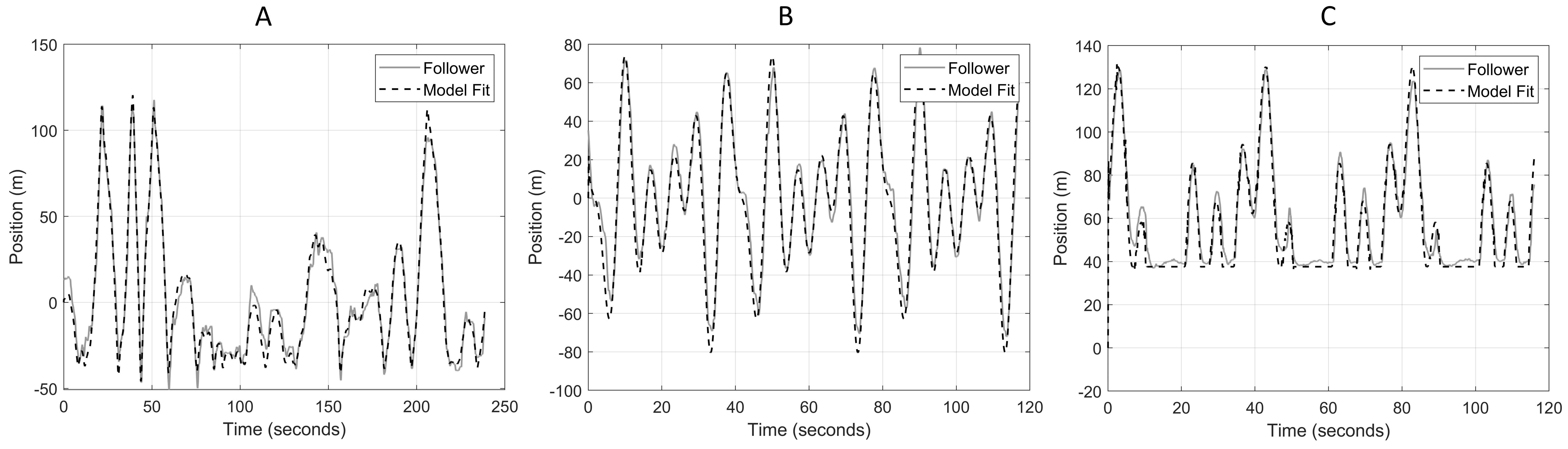}
    \caption{Fitted model versus measured output for a typical (A) white noise trajectory, (B) Fourier trajectory, and (C) Fourier trajectory with a virtual environment, leading to contact forces. This shows the accuracy of the model even for different types of inputs.}
    \label{fig:modelFit}
\end{figure*}

To test the follower's dynamics reliably and repeatably, no expert person was used to input a trajectory. Instead, trajectories were automatically generated on the expert side and sent to the follower. Two types of trajectories were generated to probe different aspects of the dynamics. First, a Fourier series of sinusoids with different frequencies, amplitudes, and phase shifts was used in every axis of position. The amplitudes were chosen to maximize usage of the available workspace. The second type of trajectory used low-pass filtered white noise by generating positions from a uniform distribution spanning the workspace, and applying a low-pass filter with sharp cutoff at the maximum comfortable frequency. Examples are shown in Figs. \ref{fig:modelFit} and \ref{fig:tracking}.

In both cases, the maximum frequency of approximately 0.63 Hz was based on frequency response tests for human teleoperation \cite{black2023ijcars}, to make trajectory following possible, though it required focus. The follower's ability to track the virtual probe's speed depends both on the frequency and amplitude of the signal. At 150 mm amplitude, it was found previously that the follower's frequency response crosses -3 dB at approximately 0.6 Hz \cite{black2023ijcars}.

A limited test of orientation tracking was also performed, to check if the position dynamics model can also be applied to orientation, indicating similar behavior in rotation as in position, at least for relatively small rotations. For this, the same white noise trajectory generation was used, but with rotations of $\pm50^\circ$ in every axis sampled uniformly about the upright position of the handle.

Every user was given one minute to familiarize themselves with the system before starting the tests. They then followed a Fourier series trajectory for 2 minutes, followed immediately by a filtered white noise trajectory for 4 minutes, with approximately 5 seconds rest in between. They were simply asked to follow the virtual probe as closely as possible. In total, 12 users completed the test (5 female, mean age 27.4). On a later date, 6 of the users completed the test again, to check time invariance.

To identify the follower dynamics by determining parameter values that best fit the model to the measured data, linear grey-box estimation was utilized with Adaptive subspace Gauss-Newton search, non-negativity constraints and regularization on all parameters, and a stability constraint on the system. This was performed with the MATLAB system identification toolbox.

\section{Results}
\subsection{Axis Coupling} \label{sec:axis}
If the axes are uncoupled, the 3- or 6-DOF system can be represented as a block-diagonal combination of the 1-DOF state space matrices. Conversely, if there is significant coupling, the off-block-diagonal elements of $A_3,~B_3,~C_3$ will be large.

To determine which is the case, the 1-DOF model was first fitted to the three axes individually to obtain parameter values for each axis. We then performed the same parameter estimation routine for $A_3,~B_3,~C_3$, but left the matrix elements completely free rather than enforcing the specific structure of Equation \ref{eqn:stateSpace}. In this way, the optimization was free to choose any state space model that constitutes a stable system. The parameter values from the individual axes were used as initial values for the optimization. Running the optimization on each subject's trajectory data, we find in Table \ref{tab:matElems} that the matrix elements we expected to be zero from the modeling indeed remain close to zero for position tracking, with errors of $<1\%$ for $A$ and $B$, and approximately $5\%$ for $C$. For orientation tracking, the same is true, but with slightly higher errors in $B$ and $C$ of $4.4\%$ and $8.3\%$, respectively. 

\begin{table}[h]
    \centering
    \caption{Mean $\pm$ standard deviation of the optimal, unconstrained matrix elements that are expected from the model to be zero, and those that are expected to be non-zero. $A_3,~B_3,~C_3$ are for position and $A_\theta,~B_\theta,~C_\theta$ are for orientation}
    \begin{tabular}{|c|c|c|}\hline
        Matrix & Expected Non-zero Elements & Expected Zero Elements \\\hline
        $A_3$ & $270.70\pm397.27$ & $0.52\pm 1.50$\\
        $B_3$ & $691.24\pm367.25$ & $11.01\pm36.89$\\
        $C_3$ & $1.41\pm0.66$ & $0.092\pm0.24$\\\hline
        $A_\theta$ & $299.91\pm408.03$ & $2.49\pm5.13$\\
        $B_\theta$ & $681.29\pm453.07$ & $30.23\pm89.64$\\
        $C_\theta$ & $0.93\pm0.092$ & $0.078\pm0.27$\\\hline
    \end{tabular}
    \label{tab:matElems}
\end{table}

We also performed a test in which the input trajectory was limited to one axis at a time, and after testing the three axes, the same set of trajectories was run again, this time with the three axes at once. It was found that the follower performance was better when following only one axis at a time. With one axis at a time, the average root mean squared (RMS) percent error for each axis $[5.27\%,~2.57\%,~5.21\%]$, while following three axes at a time led to an error of $[8.49\%,~4.11\%,~12.38\%]$ ($p<0.001$ for the error difference in every axis). However, this does not imply that the axes have any coupling. They are still uncoupled, but the model parameters change when controlling one versus three at a time.

Therefore, the follower's motions in different axes are not coupled, so we can treat each individually. Orientation tracking can use the same model, replacing mass with a moment of inertia, and the linear springs and dampers with rotational ones, at least for relatively small angles ($<60^\circ$). The full 6-DOF model is then simply a block diagonal combination of 6 copies of the above model, potentially with different parameter values. Hence, the following analysis based on Equation \ref{eqn:stateSpace} represents, without loss of generality, the three-dimensional position and orientation of the follower.
\begin{figure}[h]
    \centering
    \includegraphics[width=\linewidth]{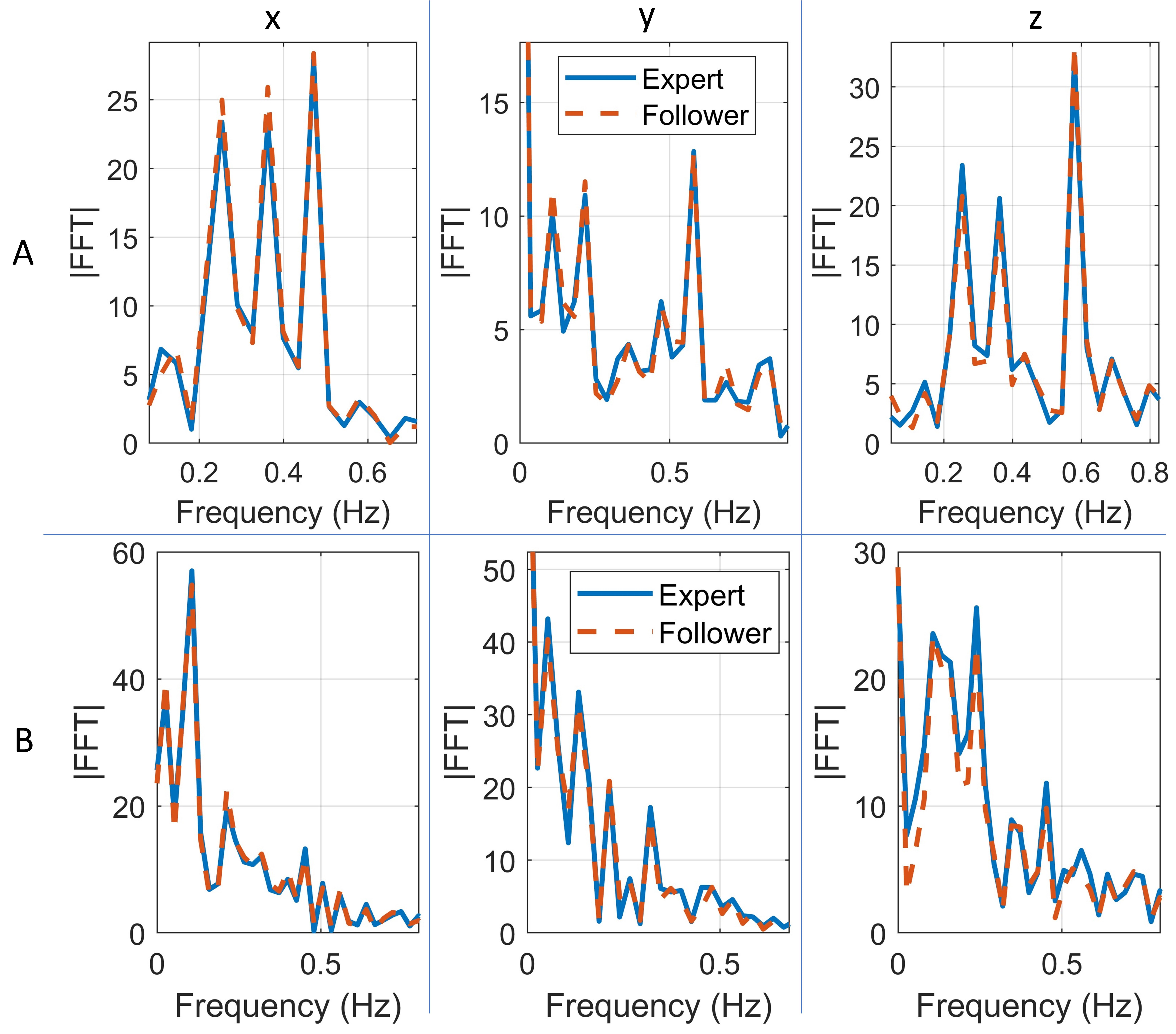}
    \caption{Fourier transforms of the Fourier series (A) and white noise (B) trajectories, showing very close agreement in frequency content between expert and follower (input and output).}
    \label{fig:fft}
\end{figure}
\begin{figure}[h]
    \centering
    \includegraphics[width=0.8\linewidth]{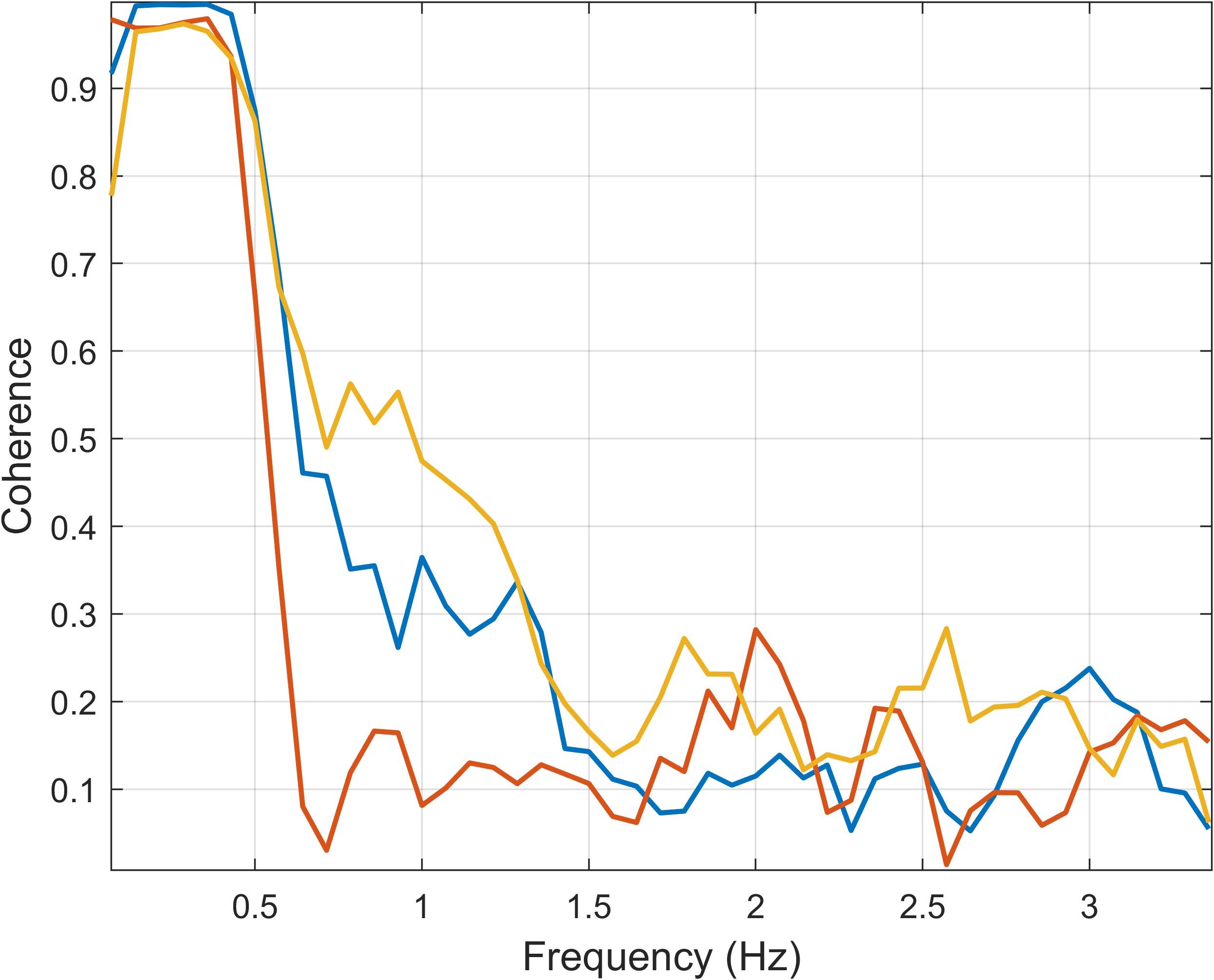}
    \caption{Coherence function of a typical follower, showing high coherence between the input and output signals of the follower up to the cutoff frequency of 0.6 Hz.}
    \label{fig:coherence}
\end{figure}
\subsection{Linearity} \label{sec:linear}
To test linearity, the first question is how well the linear model described above represents the data. The model was fitted to the data using the grey-box estimation described above. The mean of the RMS $\%$ error by axis for all subjects is $[4.42\pm1.38\%,~5.42\pm1.69\%,~6.44\pm2.15\%]$, indicating a good fit and suggesting linearity. This is shown in Fig.~\ref{fig:modelFit}. In orientation, the fit errors are slightly higher, but the predicted response is still representative: $[7.71\pm2.24\%,~4.49\pm1.86\%,~8.50\pm1.96\%]$. We therefore assume the remaining analysis applies similarly to orientation tracking.

As shown in Fig. \ref{fig:fft}, the frequency content of the follower's input and output signals are also very similar. A nonlinear system would tend to introduce new frequencies or harmonics, but this is not the case with the follower. For both filtered white noise and Fourier trajectories, the mean normalized cross-correlation by axis between the Fourier transforms of the input and output trajectories for all subjects is $[0.993\pm0.007,~0.998\pm0.001,~0.988\pm0.01]$. Similarly, the coherence function of Fig. \ref{fig:coherence} shows that the output of the follower closely matches the input before decreasing rapidly past the cutoff frequency of the signal. This also suggests linear behavior, at least up to the cutoff frequency.

\begin{figure}[h]
    \centering
    \includegraphics[width=0.7\linewidth]{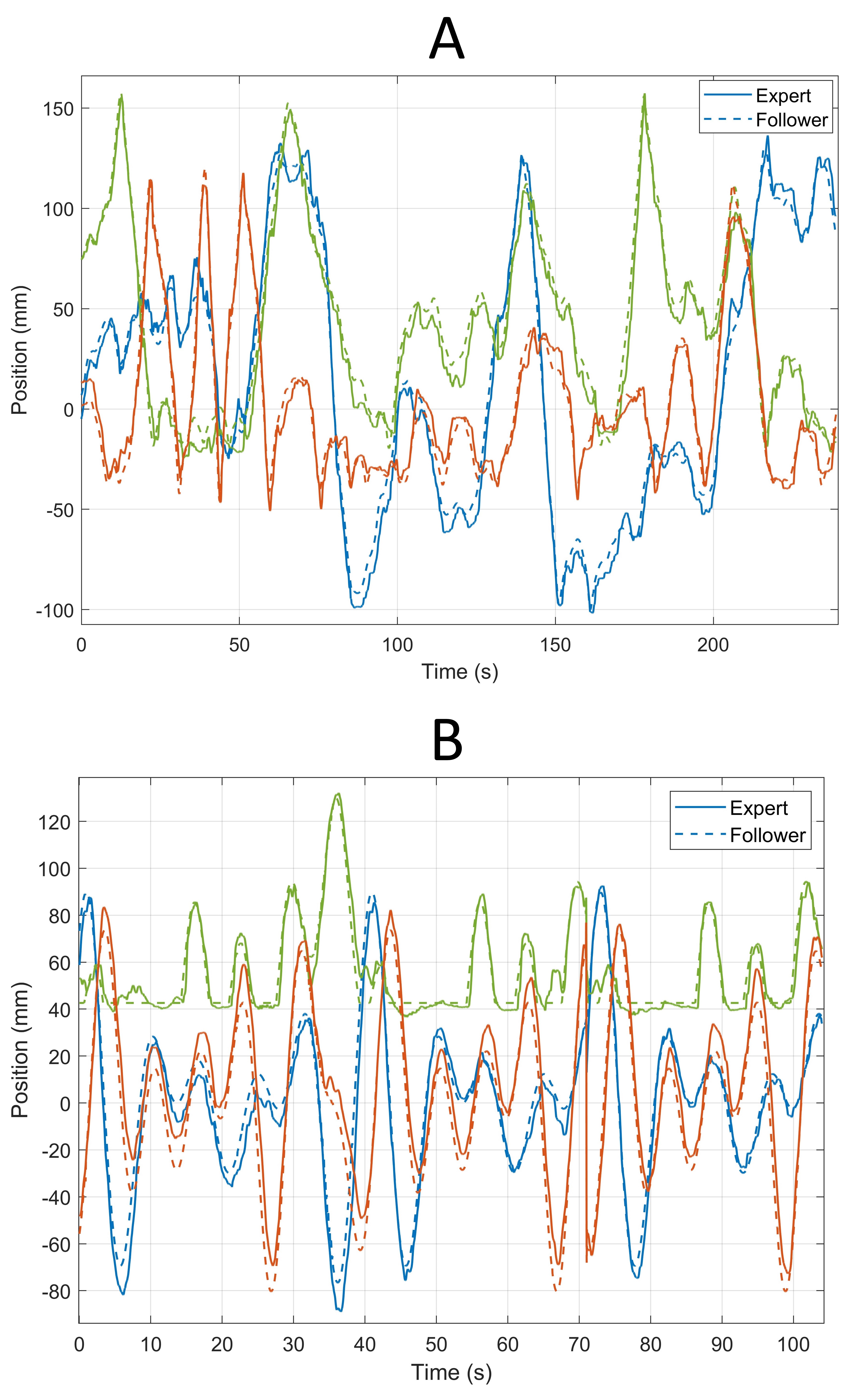}
    \caption{Input and output trajectories of the follower person, showing accurate tracking. (A) is filtered white noise while (B) is a Fourier series with a virtual environment.}
    \label{fig:tracking}
\end{figure}

Finally, we measured the correlation of the output with the filtered white noise input. It was found that the mean normalized cross-correlation by axis for all subjects is $[0.993\pm0.003,~0.987\pm0.008,~0.987\pm0.004]$, all of which are very close to unity. Thus, the correlation is very high, again suggesting that the follower dynamics are approximately linear. Fig. \ref{fig:tracking} shows the accuracy of the follower's tracking.

\subsection{Time Invariance} \label{sec:ti}
The approximate time invariance of human followers was evaluated in two ways. First, we considered whether the follower becomes fatigued during the course of a procedure and the dynamics thus change, with degrading performance as a function of time. While typical ultrasound exams are generally between 10 and 60 minutes, the motions are far slower and smaller, are limited to a surface (the patient's skin), and are interspersed with long periods of holding the probe stationary while acquiring or annotating images. Moreover, the sonographer must reposition the patient, apply ultrasound gel, and more. To determine how much the sonographer actually moves during a typical scan, we analyzed data of 11 simplified abdominal ultrasound exams from a previous study \cite{yeung2024}. In this study, the position and orientation of the ultrasound probe were tracked during the whole scan. On average, the cumulative distance moved by the ultrasound probe was $4.87\pm1.93$ m. The experiments described here consisted of 6 minutes of nearly continuous motions, leading to an average cumulative distance traveled of $6.58\pm2.92$ m. Therefore, the level of fatigue from our tests should be similar to that of a clinical ultrasound exam. 

To check for performance degradation, the 6 minutes were split equally into 2 minute sections representing the beginning, middle, and end of the scans. Model parameters were fit to each section separately, and the values were compared. Additionally, the performance of the trajectory following was evaluated during each time segment, and the accuracy of a model fitted to the beginning segment was evaluated on the middle and end segments.

It was found that the RMS tracking error changes from beginning to middle and from middle to end were $1.74\pm5.30$ mm and $-1.43\pm 6.44$ mm, respectively, and we failed to determine a statistically significant difference, though the sample size is relatively small. The model parameters of a given subject differed on average by $0.56\pm3.19\%$ between the three segments. Similarly, the RMS error of a model fitted to one time segment varied on average by $1.14\pm 4.47\%$ on the other two segments across all subjects. From these results we can conclude that the follower person is approximately time invariant over the duration of a typical ultrasound scan in normal conditions.

Second, we tested if the follower's dynamics differ on different days or times of day. This assumes, as stated in the Methods, that the person is in a normal state at both times. To answer this question, we simply repeated the original tests on a subsequent day. The model was fitted to the first set of data, and then evaluated on the data from both days. On average, the RMS error of the model compared to the measured motion increased by $0.49\pm1.19\%$ on the unseen data, showing that the dynamics of the follower did not change meaningfully. 
\begin{figure}[h]
    \centering
    \includegraphics[width=\linewidth]{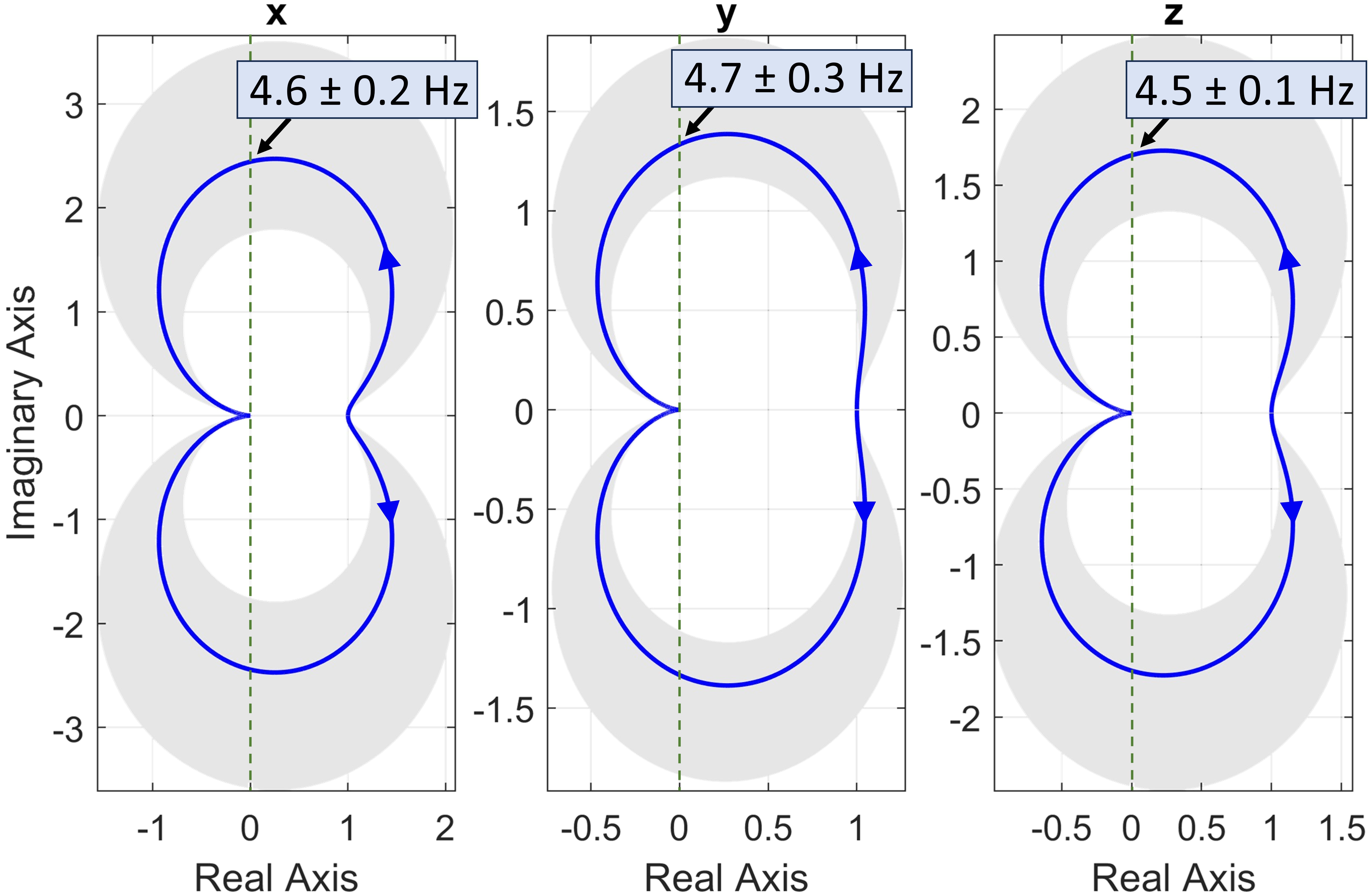}
    \caption{Nyquist plots of the three positional axes, showing the mean and range of the models fitted to all subjects. The frequencies of crossing the imaginary axis, beyond which passivity is lost, are indicated.}
    \label{fig:nyquist}
\end{figure}

\subsection{Passivity}
The passivity of the follower person in human teleoperation can be explored in several ways. First, we can use the model in Equation \ref{eqn:stateSpace}. Given that the follower model consists only of passive elements (mass, spring, and damper), their response to the input velocity should be passive. However, the input may be active. A passive system has a Nyquist plot in the right half plane \cite{jazayeri2015}. The Nyquist plots of the average model fitted to the data from the test subjects are shown in Fig. \ref{fig:nyquist}. These indicate that the follower exhibits passive behavior up to an input frequency of approximately 4.6 Hz. Note that these models are fitted to input data whose bandwidth does not reach 4.6 Hz, so this is an extrapolation. Indeed, it is difficult for the follower to achieve motions of 4.6 Hz, even at low amplitudes, so the model may become less accurate \cite{black2023ijcars}. This frequency is also much higher than would ever be necessary for a tele-guided manual task like an ultrasound exam. As a result, the loss of passivity beyond 4.6 Hz is not relevant to human teleoperation, and the follower can be considered passive for a realistic, practical frequency range.

\begin{figure}[h]
    \centering
    \includegraphics[width=0.9\linewidth]{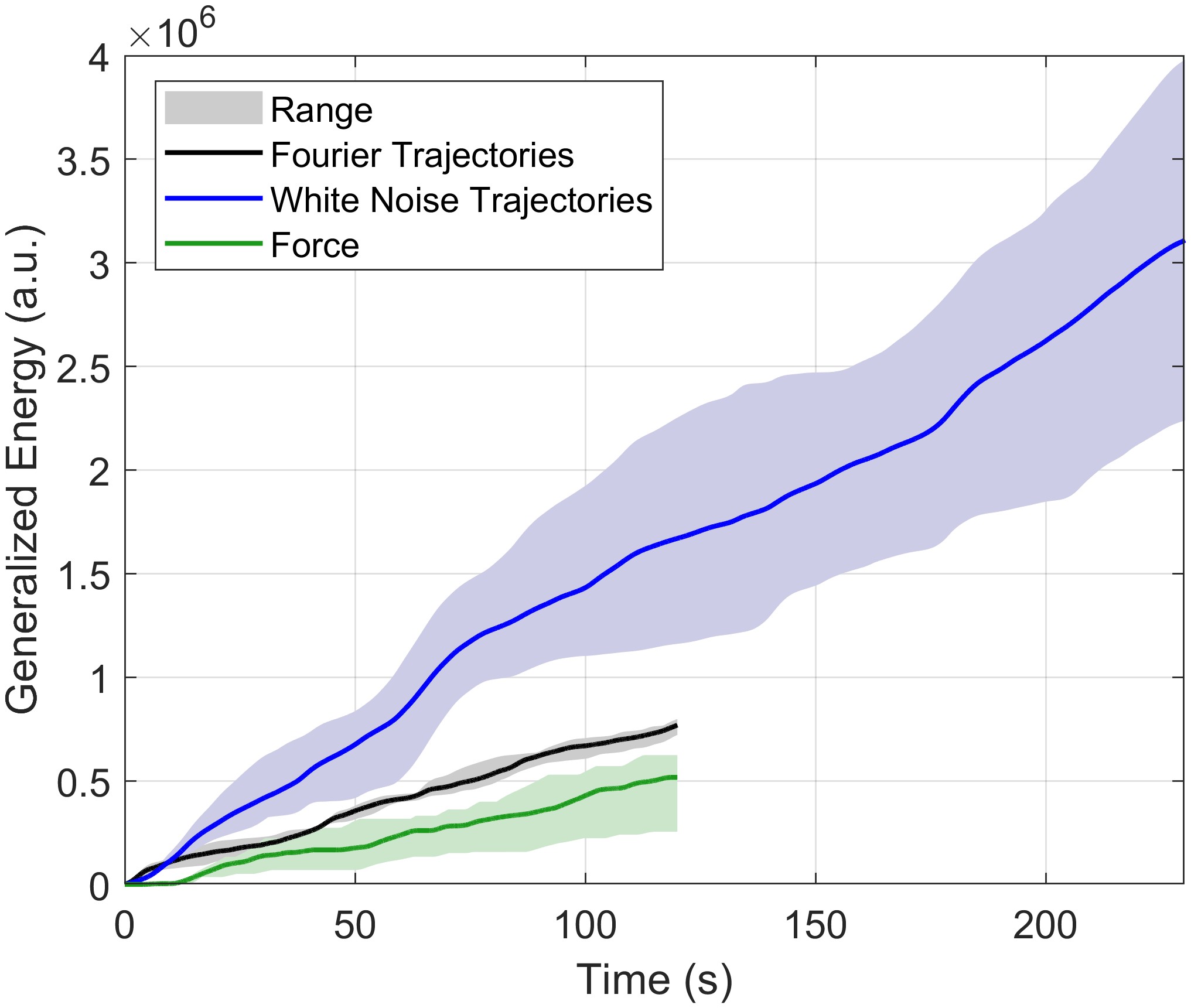}
    \caption{Mean energy over time of the filtered white noise and Fourier series trajectories, as well as the range of values from all subjects. A large excess of passivity is apparent in both cases for all users. The energy if considering the desired and measured forces is shown as well.}
    \label{fig:energy}
\end{figure}

Looking directly at the experimental data of the users, it is possible to track passivity by comparing the input and output signals of the follower. Since in the follower model, the applied force does not affect the follower's dynamics, we first replace the second row of $C$ with $\bmat{0&1}$ to output velocity instead. Now, the energy at measurement sample $K$ is approximately
\begin{equation}
    E[k] = \Delta T\sum_{j=0}^k\pmb{u}[j]^\top\pmb{y}_f[j]
\end{equation}
This is plotted for every subject in Fig. \ref{fig:energy}, showing that the energy is increasingly positive with time, displaying a large excess of passivity for both types of trajectories, for every subject. The energy including the input and output force (i.e. with $C$ as in Equation \ref{eqn:stateSpace}) is also plotted and shows passive behavior as well.

\subsection{Stochasticity}
\begin{figure}[h]
    \centering
    \includegraphics[width=0.8\linewidth]{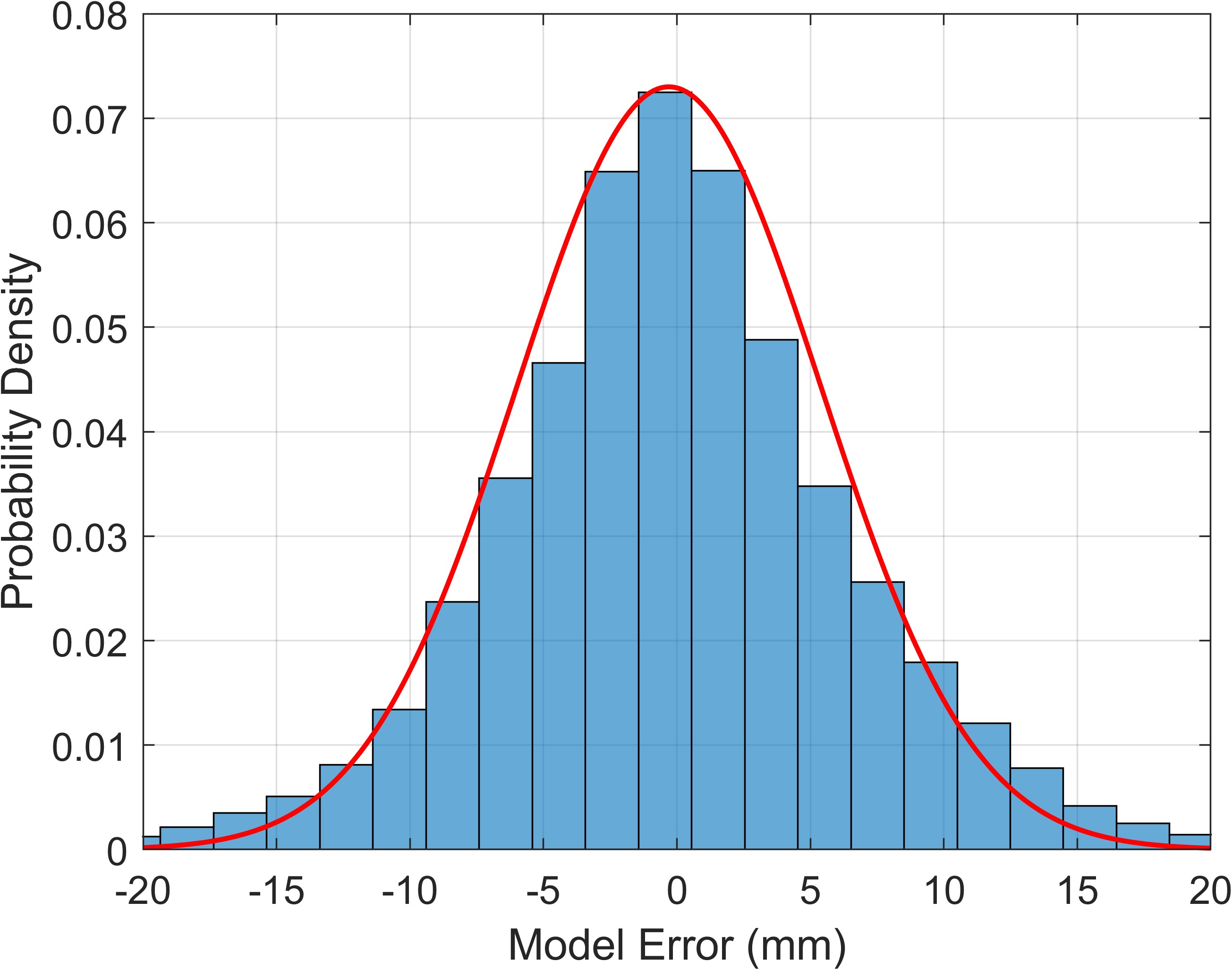}
    \caption{Error distribution of the deterministic models (Equation \ref{eqn:stateSpace}) compared to the measured trajectories, showing that the follower's motion is normally distributed about the expected trajectory.}
    \label{fig:errorHist}
\end{figure}

All of these models and conclusions of linearity, time-invariance, and passivity are approximate. If the same follower were asked to track the same trajectory 100 times, they would certainly never carry out precisely the same sequence of motions. Instead, the follower acts like the model above plus some additive noise. To characterize the noise, the fitted models were subtracted from the measured signals, and the resulting error values from all samples of all the subjects were plotted in a histogram in Fig. \ref{fig:errorHist}. The error values fit well onto a Gaussian distribution with zero mean and $\sigma=7.00$ mm. The Bhattacharyya distance between this distribution and the experimental data is $1.24e-6$, while the distance between two sets of data generated from the same distribution is $7.47e-7$. 

Therefore, a  more accurate follower model, taking into account the person's stochasticity, is as follows
\begin{align}\label{eqn:stateSpace}
    \pmb{\dot{x}}_f &= \bmat{\dot{x}_f\\\ddot{x}_f}=A\pmb{x}_f+B\pmb{u}\nonumber\\
    \pmb{y}_f &= \bmat{x_f\\f_f}=C\pmb{x}_f + \bmat{\mathcal{N}(\mu=0,\sigma=0.007)\\f_{rand}}
\end{align}
Where $f_{rand}$
\begin{equation}
    f_{rand} = \begin{cases}
        \mathcal{N}(\mu=0,\sigma=0.007k_p) & \text{During contact}\\
        0 & \text{Otherwise}
    \end{cases}
\end{equation}
The $\sigma$ are in units of meters and Newtons, and $k_p$ is the stiffness of the patient tissue. Note that the randomness in this model affects only the output and not the internal state. It achieves the desired effect, but an even more realistic model might apply the random variation to the follower's accelerations. However, it is difficult to comment on their distribution from these experiments as the double derivative of the measurements leads to noisy acceleration data.

\section{Discussion and Conclusion}
This paper has shown through modeling and empirical evidence that for the purpose of practical controller design and analysis, the follower in a human teleoperation system can be treated as a passive LTI system with minimal inter-axis coupling. The latter allows analysis to be simplified to 1 DOF at a time, while the LTI nature permits application of the vast field of linear control. The passivity of the human follower enables use of methods to passivate the communication channel, such as wave variables \cite{niemeyer1991,aziminejad2008} and time domain passivity \cite{ryu2004}.

One limitation of this study is the relatively small sample size. However, the results are very clear and correspond well with the modeling. A further limitation is that the data was collected using trajectories of large amplitude and correspondingly low bandwidth so the follower could track them. The large amplitude maximizes the position signal but means that the fitted models are based on only low frequency data and may not generalize as well to higher frequencies. In practice, the follower cannot achieve much higher frequencies because of limitations in perception, processing, and motion speed, so this limitation likely does not affect the validity of the results for human teleoperation. However, future work could validate the results by testing lower amplitude, higher frequency trajectories as well.

The presented analysis applies in ``normal" conditions, and will become less accurate in extreme situations. For example, the follower response will saturate at larger force values when in contact with a patient, to avoid injuring the patient. Furthermore, the Nyquist diagram shows that the follower acts as a low-pass filter, as expected, with decreasing gain and increasing phase shift as the frequency grows. However, in practice, beyond a certain input speed, the follower becomes unable to perceive and react to the input, so the model ceases to be representative. It thus breaks down for high-frequency or large-amplitude signals, when the follower can no longer track the motion well. If the follower is mentally or physically affected during the procedure, their performance will also change.

Although a given person is approximately LTI, as demonstrated above, and shares approximately the same model structure as other people, the model parameters are unique to the individual. In analogy to a typical control system such as an inverted pendulum, each follower has different pendulum mass and length. Thus, while the same control structure may be effective for most users, the optimal parameters may vary between individuals as they would between pendulums of varying dimensions. Consequently, adaptive controllers should be explored that can tune themselves according to the identified dynamics. 

This paper has also demonstrated a stochastic element to the follower model that is less prominent in robotic systems. How the variance of the stochastic part depends on factors such as follower fatigue or focus should be studied in future and may enable development of controllers that are robust to human factors. Stochastic and robust control approaches may also be relevant.

\bibliographystyle{ieeetr}
\bibliography{refs}
\end{document}